\documentclass{aa} 

\usepackage{graphicx}
\usepackage{txfonts}
\usepackage{hyperref}
\usepackage{float}

\begin{document} 

   \title{3D structures of the base of small-scale recurrent jets revealed by Solar orbiter}

   \author{Xiaohong Li\inst{1} 
   \and
   Sami K. Solanki\inst{1} 
   \and
   Thomas Wiegelmann \inst{1} 
   \and
   Gherardo Valori\inst{1} 
   \and 
   Daniele Calchetti\inst{1} 
    \and
   Johann Hirzberger\inst{1} 
   \and
   Juan Sebasti\'an Castellanos Dur\'an \inst{1} 
   \and  
   Joachim Woch\inst{1}
   \and
   Achim Gandorfer\inst{1}
     \and
   the Solar Orbiter team
   }
   \institute{Max Planck Institute for Solar System Research, Göttingen D-37077, Germany \\
    \email{lixiaohong@mps.mpg.de}
              }          
              
   \date{Received ; accepted }

  \abstract
   {Solar jets, characterized by small-scale plasma ejections along open magnetic field lines or the legs of large-scale coronal loops, play a crucial role in the dynamics of the solar atmosphere. They are often associated with other solar active phenomena, including campfires, filament eruptions, coronal bright points, flares, and coronal mass ejections.}
   {Although spectral and EUV images have been widely used to analyze the formation and evolution of jets, the detailed three-dimensional structure at the base of the jet has not been studied in detail due to the limitations in the spatial resolution of observations.}
   {Solar Orbiter enables us to investigate the structure of solar jets with much higher spatial and temporal resolutions and from a different angle than from Earth. By combining observations made by instruments onboard Solar Orbiter with data from the Solar Dynamics Observatory, we analyzed recurrent solar jets originating in a mixed-polarity region near an active region. Additionally, we employed potential field and magneto-hydro-static extrapolation techniques to determine the magnetic field topology associated with the jets.}
   {The jets display dynamic, multi‐strand outflows emanating from compact bright kernels above the magnetic inversion line, with apparent speeds exceeding 100 km s$^{-1}$. Magnetic field evolution reveals continuous flux cancellation at the jet footpoints. Throughout the sequence, base flows are confined within quasi‐separatrix layers, with the highest velocities and temperatures located near coronal null points. Over four eruptions, the magnetic topology evolves from a simple fan–spine configuration with a single null to a more complex dome‐shaped base containing multiple nulls with separatrix curtain, accompanied by a morphological transition from narrow, well‐collimated spire to broader, fragmented outflows.} 
   {These results provide the first direct observational evidence that dynamic changes in null‐point geometry modulate jet morphology and energetics via successive reconnection episodes.}


   \maketitle

\section{Introduction}  \label{sec:intro}

Solar jets are small-scale dynamic plasma eruptions driven by photospheric motions and ejected into the corona, often guided by open magnetic field lines or the legs of large-scale coronal loops. These jets exhibit multi-thermal structures, allowing them to be observed across a broad range of wavelengths, including white light when observed coronagraphically, X-rays, ultraviolet (UV), and H-alpha \citep{Yokoyama1995, Chae1999, Shibata2007, Cirtain2007, Sterling2015}. Their multi-wavelength visibility underscores the intricate thermal and magnetic processes involved in their formation and evolution.

Solar jets occur in diverse solar environments and span a wide range of temporal and spatial scales \citep{Bharti2017, Tian2018}. Their lengths can range from a few megameters to several solar radii, particularly when observed in white light \citep{Moore2010, Raouafi2016}. With advancements in solar observations, such as the launch of the Solar Orbiter \citep{Garcia2021, Muller2020}, the detection of smaller-scale jets has become possible in EUV radiation, revealing details previously unresolved. These findings highlight the ubiquitous nature of jets, occurring in active regions, coronal holes, and quiet Sun areas \citep[e.g.,][]{Chitta2023, Chitta2025, Shi2024}.

Observationally, solar jets are characterized by two main components: the base and the spire. The base is typically bright and localized, marking the region of magnetic reconnection, where stored magnetic energy is rapidly released, heating and accelerating plasma. This bright base often shows dynamic features such as swirling motions or compact brightenings, thought to be associated with flux cancellation, emerging flux, or sheared magnetic fields \citep{Chae2004, Panesar2016}. Magnetic reconnection occurring in these bases produces bi-directional outflows: plasma is propelled upward along magnetic field lines, forming the spire, while downward-directed plasma contributes to heating at the base \citep{Moreno2013, Hardi2014, LiX2018, Pontin2022}.

The spire, on the other hand, is an elongated and collimated structure extending radially into the corona. It consists of hot plasma streams visible in X-rays and UV, as well as cooler plasma observed in H-alpha \citep{Nistico2009, Chandrashekhar2014}. Observations have revealed fine-scale structures within spires, such as strands or helical patterns, which may reflect the underlying magnetic topology and the plasma dynamics \citep{Pariat2015, Shen2021}. The spire's motion often includes oscillations, twisting, and rotational components, suggesting the presence of Alfvén waves or torsional motions \citep{Cirtain2007, Sterling2015}.

Advancements in observational capabilities have deepened our understanding of jet structures. For instance, Solar Dynamics Observatory (SDO, \citealt{Pesnell2012}) has facilitated the statistical analysis of jets, linking their occurrence to underlying photospheric magnetic activity such as flux emergence or cancellation \citep{Panesar2016, Joshi2024}. High-resolution observations reveal features like fine-scale filaments embedded inside the jet base arch \citep{Sterling2015, LiX2015, Shen2021} and the plasmoids ejected along the jet spire \citep{LiX2018, LiX2019, Zhang2019}. Recent findings also highlight the role of chromospheric processes, such as plasma evaporation and cooling, in shaping the observed multi-thermal characteristics of jets \citep{Schmieder2022, LiX2023}.

Despite significant progress, some aspects of jet structures remain poorly understood, particularly the fine-scale dynamics and morphology of their bases. For instance, while the spire is often the focus of observational studies, the jet base records where energy released by the reconnection is deposited, driving intense heating and launching accelerated plasma. However, the brightness of the base often obscures finer details, limiting our understanding of the processes governing jet initiation and evolution. Recent studies suggest that the magnetic topology at the base, such as null points, separatrix surfaces, or quasi-separatrix layers (QSLs), plays a critical role in determining jet morphology and dynamics \citep{Pariat2015, Pontin2022}.

Solar Orbiter, with its high spatial and temporal resolution and its capability to observe the Sun from varying angles, provides unprecedented opportunities to investigate jets. It has enabled the detection of smaller-scale jets and allowed multi-perspective observations that help constrain the three-dimensional structure of jet bases and spires. These insights are crucial for advancing our understanding of the physical mechanisms driving these ubiquitous solar phenomena, particularly their role in the transport of mass and energy in the solar atmosphere. In this paper, we exploit the unique opportunities offered by Solar Orbiter’s high spatial and temporal resolution and its varying viewing angles, together with \emph{SDO} at a $45^\circ$ separation, to present a detailed 3D study of recurrent jets focusing in particular on the complex dynamics and multi-thermal structures at their bases. In Sect. \ref{sec:obs}, we give an overview of the observations. In Sect. \ref{sec:res}, we present the dynamics and the magnetic structures of the jet. The conclusion and discussion are presented in Sect. \ref{sec:con}.

  \begin{figure*}[!htbp]
  \centering
  \includegraphics[trim=2cm 1.5cm 2cm 1cm, clip, width=\textwidth] {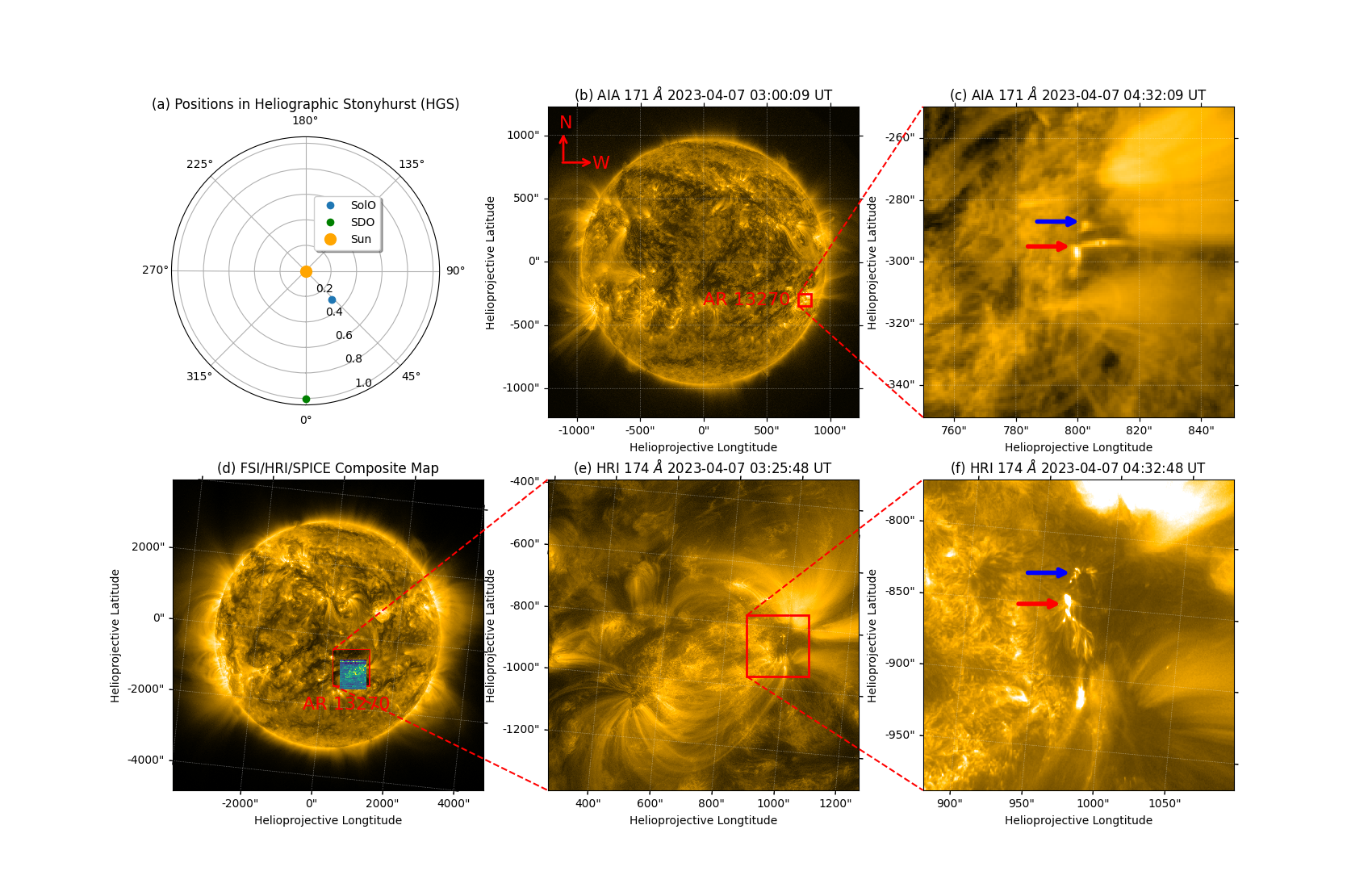}
     \caption{Overview of the recurrent jet events on 7 April 2023. (a) Relative positions of the Sun, SDO, and Solar Orbiter. (b) Full-disk image of the Sun taken by SDO/AIA at 171 {\AA}. (c) SDO/AIA 171 {\AA} image of a sub-region in AR 13270. (d) Composite map combining the full-disk image of SO/EUI-FSI at 174 {\AA}, the SO/EUI-HRI 174 \AA{} image, and the SO/SPICE ``O V 760.2 ... Ne VIII 770 (Merged)'' spectral window. (e) SO/EUI-HRI 174 {\AA} image focusing on AR 13270. (f) Zoom into the recurrent jets observed in AR 13270 as seen by SO/EUI-HRI. The blue and red arrows in panels (c) and (f) indicate two distinct jet structures.}
        \label{Fig1}
  \end{figure*}

\section{Observations}  \label{sec:obs}

We use data from the Solar Orbiter and SDO obtained on April 7, 2023, covering approximately four hours. The Solar Orbiter observations were conducted under the Solar Orbiter Observing Plan (SOOP): R\_SMALL\_HRES\_HCAD\_RS-burst. Figure \ref{Fig1}(a) shows the relative positions of SDO and Solar Orbiter on that day. The Solar Orbiter was positioned at a distance of about 0.3 AU from the Sun, with a separation angle of 45$^\circ$ relative to SDO. 

As the Solar Orbiter was near perihelion, it provided very high-resolution data. We utilized level-2 174 {\AA} data from the High Resolution Imager (HRI) of the Extreme Ultraviolet Imager (EUI; \citealt{Rochus2020}). The field of view (FOV) was 2048 $\times$ 2048 pixels, with a spatial resolution of about 108 km per pixel and a time cadence of 3 s. We also employed one 174 {\AA} dataset from the Full Sun Imager (FSI). The High Resolution Telescope (HRT; \citealt{Gandorfer2018}) of the Polarimetric and Helioseismic Imager (PHI; \citealt{Solanki2020}) onboard Solar Orbiter observed the full Stokes profiles of the Fe I 6173 {\AA} line at five wavelength positions in the line and a nearby continuum point. The reduction and calibration of SO/PHI-HRT data were performed following the methods described in \citet{Kahil2022, Sinjan2023, Bailen2024}. The SO/PHI-HRT data have a spatial resolution of 0.5$\arcsec$ per pixel (approximately 110 km per pixel at 0.3 AU at disk center). 

We also used the SPectral Imaging of the Coronal Environment (SPICE; \citealt{SPICE2020, Fludra2021}) level-2 data obtained during this SOOP, which included 96-step rasters with an exposure time of 10 seconds. The slit was 4$\arcsec$ wide, and the FOV is shown in Figure \ref{Fig1}(d). The SO/SPICE data consisted of 834 pixels along the slit, and the pixel scale along the slit was 1.098$\arcsec$ (approximately 242 km on the Sun). Among the available spectral lines, we selected the following: O II at 718.5 {\AA} ($log T [K]$ = 4.7), C III at 977.03 {\AA} ($log T [K]$ = 4.8), O III at 703.8 {\AA} ($log T [K]$ = 5.0), O V at 760.4 {\AA} ($log T [K]$ = 5.4), O VI at 1031.93 {\AA} ($log T [K]$ = 5.5). 

The Atmospheric Imaging Assembly (AIA; \citealt{Lemen2012}) on board SDO provided full-disk EUV and UV images with a pixel size of 0.6$\arcsec$ ($\sim$435\,km). In this study, we use data from the AIA 94, 171, and 304 {\AA} channels with a cadence of 12 s, which have strong responses at logarithmic temperatures of approximately 6.8, 5.8, and 4.7, respectively. Due to the relative difference in distance of Solar Orbiter and SDO from the Sun, a difference of 380 seconds in the light travel time from the Sun to the two spacecraft was present. In this study, observation times of the Solar Orbiter datasets have been adjusted to align with SDO.

On April 7, 2023, active region (AR) 13270 was observed near the limb by SDO and near the disk center by Solar Orbiter, as shown in Figures \ref{Fig1}(b) and \ref{Fig1}(d). Numerous jets were detected within this AR, recurring multiple times between 03:00 and 07:00 UT. The stereoscopic perspectives provided by SDO and Solar Orbiter allowed us to investigate these jets both from the side (as observed by SDO) and from above (as observed by Solar Orbiter). Figure \ref{Fig1}(c) reveals arch-like base structures and elongated spires of the jets as seen by SDO, while the high-resolution SO/EUI-HRI images from Solar Orbiter captured the detailed structures of the jet base, as illustrated in Figure \ref{Fig1}(f). As part of the online material, \texttt{Movie1.mov} shows the temporal evolution of the jets as seen by SDO/AIA 304 {\AA} and SO/EUI-HRI 171 {\AA}. These complementary perspectives provide a comprehensive 3D view of the jet morphology and dynamics.

 \begin{figure*}[!htbp]
  \centering
   \includegraphics[trim=2.5cm 2.5cm 2.5cm 2.5cm, clip, width=\textwidth]{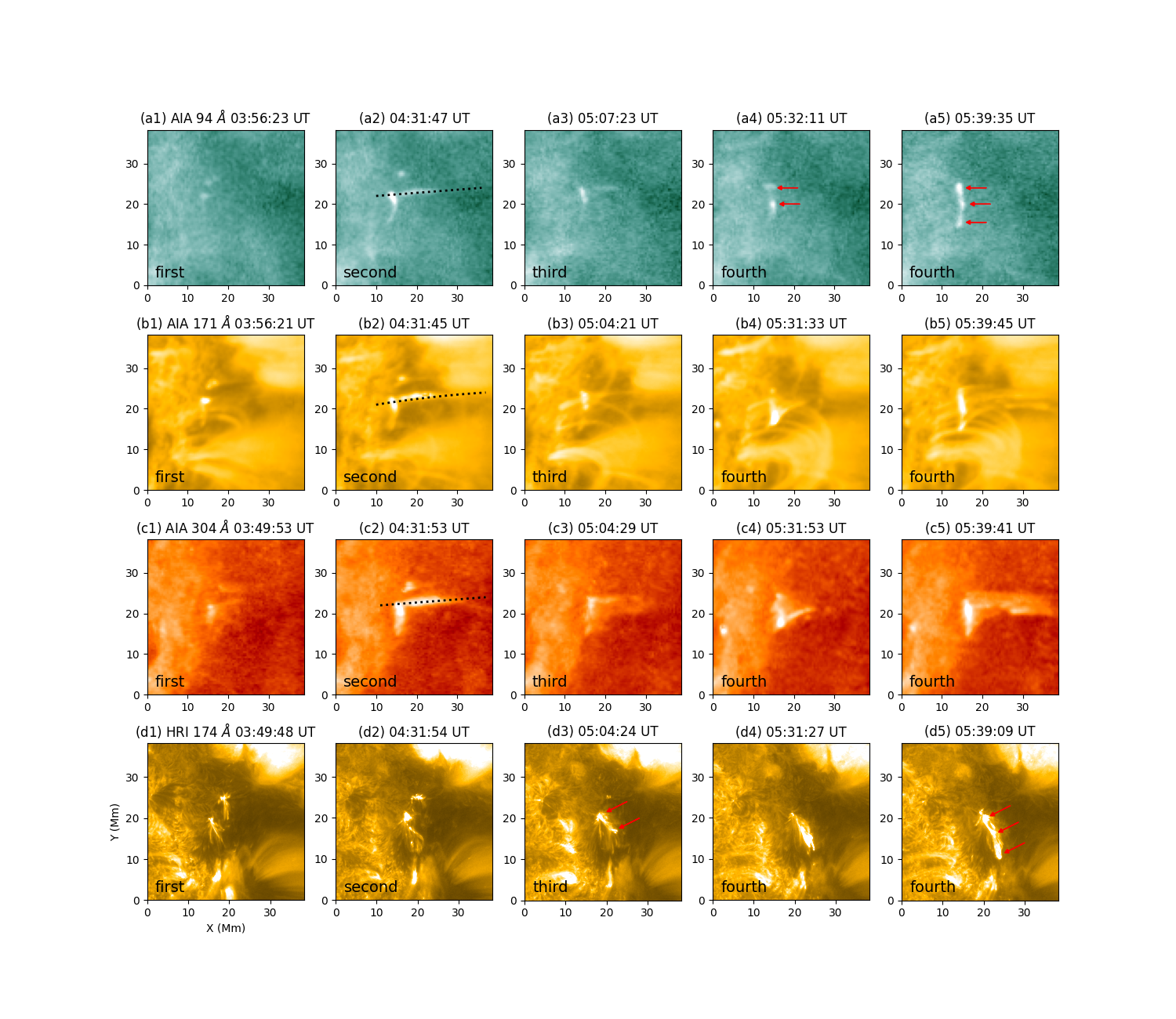}
     \caption{SDO/AIA 94 {\AA} (panels (a1$-$a5)), 171 {\AA} (panels (b1$-$b5)), 304 {\AA} (panels (c1$-$c5)) and SO/EUI-HRI 174 {\AA}  (panels (d1$-$d5)) images showing the recurrent jets during their four eruptions. The dotted lines in the second column indicate the positions where the slices are taken, and the temporal evolution of these slices is shown in Figure \ref{Fig3}. The red arrows in panels (a4), (a5), (d3) and (d5) indicate the bright kernels at the jet base.
    }
        \label{Fig2}
  \end{figure*}

  \begin{figure}[!htbp]
  \centering
  \includegraphics[width=\columnwidth]{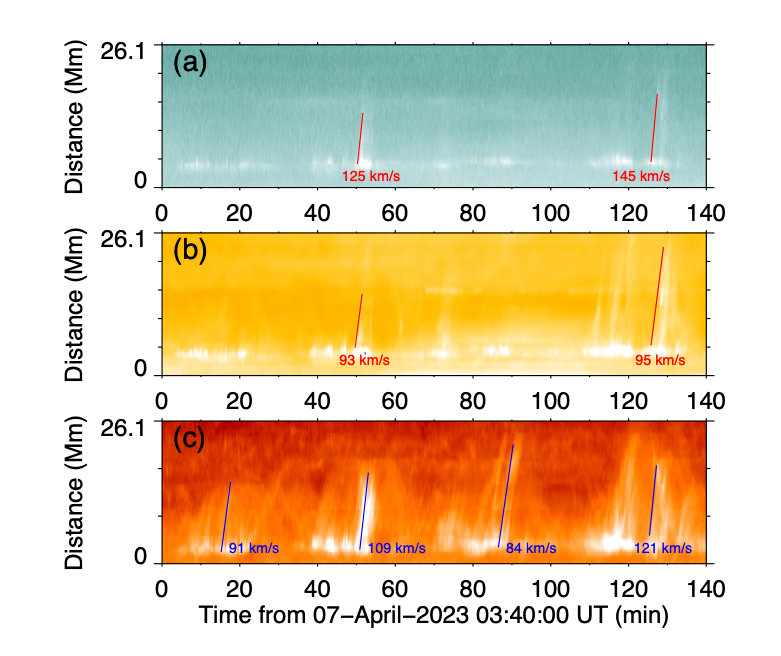}
     \caption{Time-slice plots of the SDO/AIA 94 {\AA} (panel (a)), 171 {\AA} (panel (b)), and 304 {\AA} (panel (c)) images along the spire of the southern jet marked by the dotted lines in Fig. \ref{Fig2}. The solid lines indicate the trajectories of the jet structures, with the corresponding velocities displayed alongside.}
    \label{Fig3}
  \end{figure}

\section{Analysis and Results} \label{sec:res}

\subsection{EUV observation}
Figure \ref{Fig2} illustrates four eruptions of the recurrent jets studied here, which start at around 03:45 UT, 04:20 UT, 04:50 UT and 05:30 UT, respectively. From top to bottom rows, we present SDO/AIA 94 {\AA}, 171 {\AA}, 304 {\AA}, and SO/EUI-HRI 174 {\AA} images, showcasing the jet morphology near the four eruption peak times. In the hot 94 {\AA} channel, the base area is more obvious, indicating the heating caused by magnetic reconnection, while the spire is distinctly visible in the 304 {\AA} images, suggesting the presence of cooler, denser plasma. During the first two eruptions, there are two narrow jets. In SDO observations, the jets exhibit the typical inverted-Y morphology as expected from the side view, whereas in the SO/EUI observations they appear as a bright core from which multiple bright strands extend outward in different directions. In the third eruption, the base area of the south jet expands, as two bright kernels are visible in the SO/EUI-HRI 174 {\AA} image (see the arrows in panel (d3)). Meanwhile the jet spire also becomes wider as shown in panel (c3), suggesting a more complex outflow structure than in the earlier events. During the fourth eruption, the jet starts from the south bright kernel (see the fourth column), and then develops multiple bright kernels at the base (see panels (a5) and (d5)). The spire becomes markedly broader and is composed of numerous intertwined strands extending outwards (see panels (b5) and (c5)). To investigate the eruptions in more detail, we extract time–distance plots along slits placed on the jet spires, at the positions indicated by the dotted lines in the second column of Fig. \ref{Fig2}. The corresponding time-distance plots are presented in Fig. \ref{Fig3}. The evolution along these slices clearly reveals four distinct eruptions with well-defined onset and decay phases over the course of 140 min. By tracking specific structures, we determined that the projected velocities of the jets range between 91 km s$^{-1}$ to 121 km s$^{-1}$, which is in good agreement with the typical speed range of 80–150 km s$^{-1}$ reported in previous studies \citep{Raouafi2016, Shen2021}.

\begin{figure*}[!htbp]
  \centering
  \includegraphics[trim=1cm 2.6cm 1cm 3cm, clip, width=0.95\textwidth]{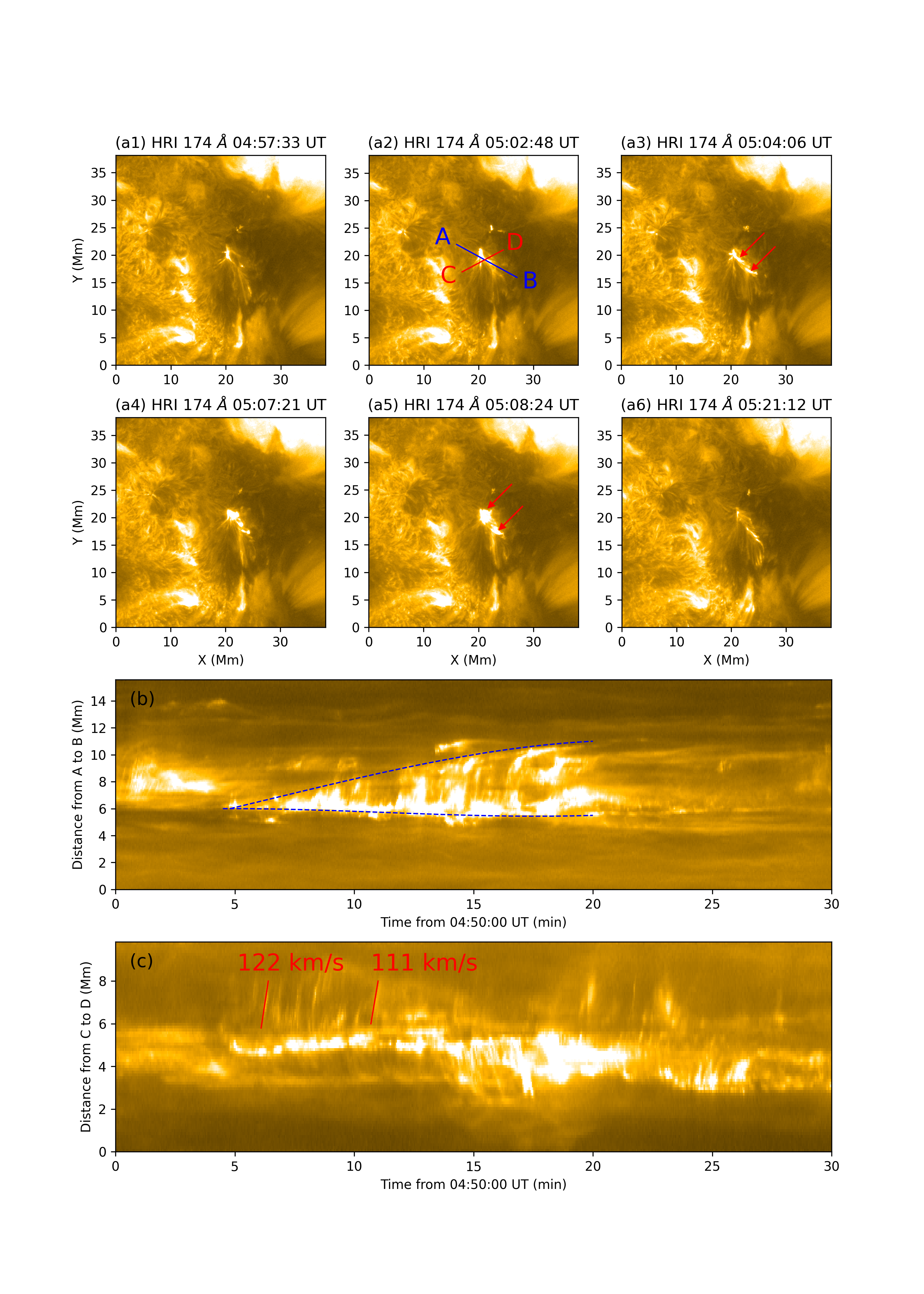}
     \caption{Panels (a1)$-$(a6): SO/EUI-HRI 174 {\AA} images showing the evolution of the third eruption of the recurrent jets. In panels (a3) and (a5), the red arrows denote two bright points. Panel (b): Temporal evolution along the position indicated by the line `A--B' in panel (a3). Panel (c): Time-distance plot along the position marked by the line `C--D' in panel (a3).}
        \label{Fig4}
  \end{figure*}

The unprecedented spatial resolution of the SO/EUI-HRI 174{\AA} observations reveals the development of the jet base in considerable detail. Since the jet morphology changes rapidly during the third eruption, we focus on this phase in Fig. \ref{Fig4}. At first, the jet appears to be similar as in previous eruptions, with one bright core and multiple bright strands extending outwards, as depicted in panel (a1). From the online \texttt{Movie1.mov} we can see flows coming out from the bright kernel. Subsequently, the jet develops southwards, and at around 05:04 UT, another bright kernel appears, as shown in panels (a3) and (a5). At this stage, the jets enter a very dynamical phase, which involves expansion of both bright kernels and enhanced outward flows.
To capture the expansion of the bright area and the outflows, we created slices in different directions, indicated by the blue and red lines in panel (a2). Fig. \ref{Fig4}(b) presents the time-slice plot along the line connecting A-B, revealing that the base area of the jet expands to approximately 5 Mm, consistent with the width of the jet base observed in the SDO/AIA images. Fig. \ref{Fig4}(c) displays the outflows along slice C-D, with the projected velocities of two distinct structures estimated to be more than 100 km s$^{-1}$. There are also flows going the other direction (near 05:04 UT). After the eruption, the brightness of the base area diminishes, revealing a multi-pronged fan structure, as shown in panel (a6) of Fig. \ref{Fig4}. Overall, the third eruption progresses from a single-kernel, narrow-strand configuration to a two-kernel, broadened, multi-strand outflow, providing a natural continuation of the morphological trend described in Fig.~\ref{Fig2}. In the subsequent fourth eruption, the activity initiates from the southern bright kernel formed in the third event, but strong flows also reappear in the northern section, further broadening the spire and reinforcing the multi-kernel, multi-strand character of the jet as demonstrated before in Figure ~\ref{Fig2}.

\subsection{Magnetic field evolution}
\begin{figure*}[!htbp]
  \centering
  \includegraphics[trim=1cm 0cm 0cm 2cm, clip, width=0.8\textwidth]{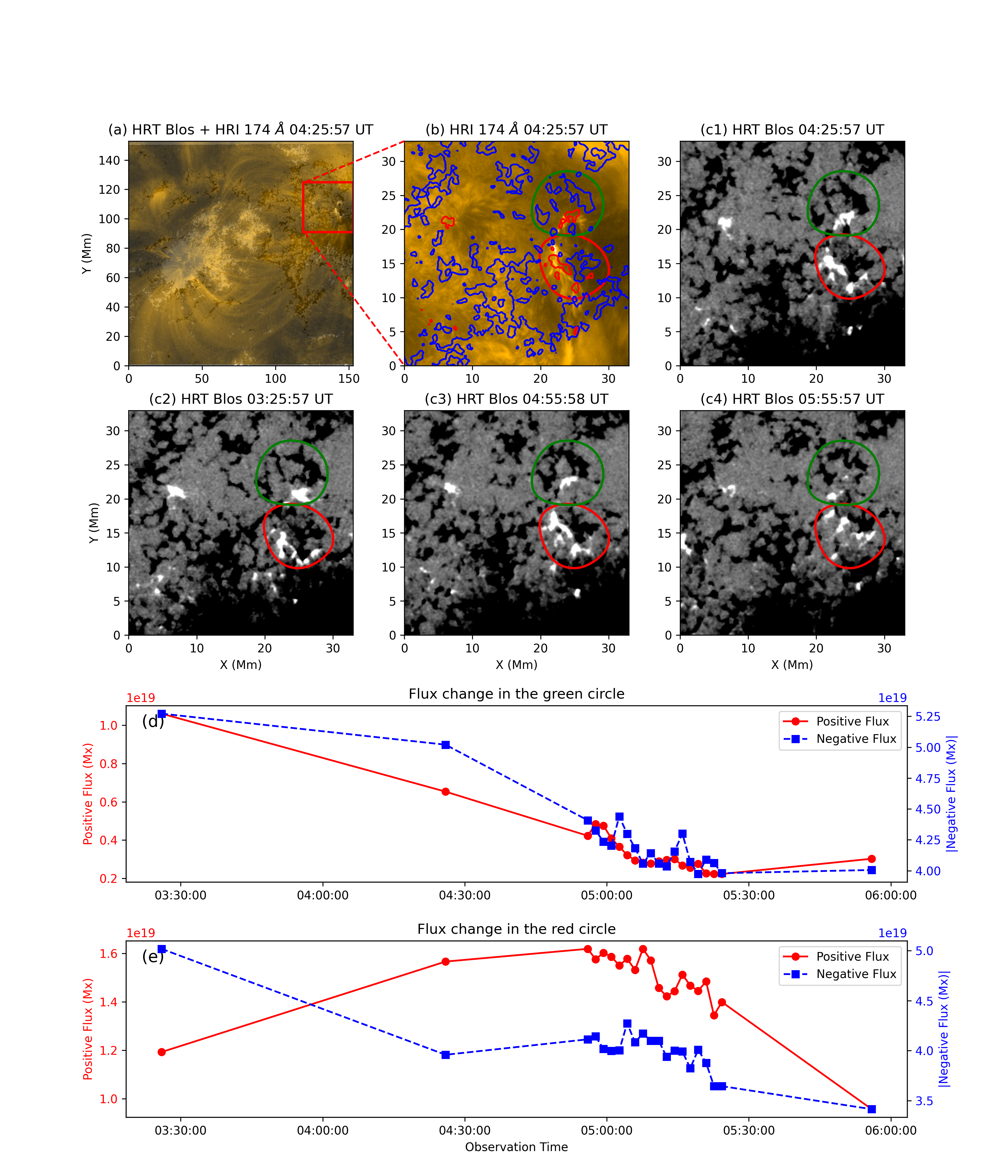}
     \caption{Evolution of the magnetic structures at the footpoints of the recurrent jets. Panel (a): SO/PHI-HRT $B_{\mathrm{LOS}}$ map overlaid with SO/EUI-HRI 174 {\AA} images of the entire active region (AR). The red square outlines the sub‐field shown in the subsequent panels where the recurrent jets occur. Panel (b): SO/EUI-HRI 174\,\AA\ image cropped to the red‐boxed region in (a). Panel (c1): Co‐spatial SO/PHI-HRT magnetogram of the same sub‐field. Panels (c2)–(c4): Temporal sequence of magnetograms showing the evolution of the magnetic field in the jet footpoint regions. The green and red circles enclose the north and south jets, respectively. These regions were chosen to remain as isolated as possible from surrounding flux concentrations. Panels (d) and (e): Time series of the magnetic flux calculated within the green and red circles, respectively, illustrating the flux changes in the two jet footpoint regions.}
    \label{Fig5}
  \end{figure*}

To investigate the evolution of the magnetic field at the jet base, we co-aligned the SO/PHI-HRT line-of-sight (LOS) magnetograms with the SO/EUI-HRI observations, as shown in Figure \ref{Fig5}(a). The SO/EUI-HRI 174 {\AA} image and SO/PHI-HRT LOS magnetogram in the red box are shown in panels (b) and (c1). In panel (b), the red and blue contours denote the +100 G and –100 G isolines of the SO/PHI-HRT LOS magnetic field, respectively. Panels (c2)$-$(c4) illustrate the evolution of the $B_{\mathrm{LOS}}$ maps at the jet base region. The footpoint region of the northern and southern jets are marked with green and red circles, respectively. These regions were chosen to enclose all the motions of the jet base, and remain as isolated as possible from surrounding flux. We can see that the jet base is situated along the magnetic inversion line, where filamentary structures are observed to connect to surrounding opposite polarities. Positive polarity magnetic flux is interacting and canceling with the nearby negative polarity fields. Eventually, both polarities weaken and decrease in magnitude. This configuration suggests that magnetic reconnection at the inversion line could play a key role in triggering the jet eruptions, consistent with the scenario proposed by \citet{Panesar2016}. 

To quantitatively analyze the magnetic cancellation process, we calculated the variation of the magnetic flux within the footpoint region of the jets. The temporal evolution of the magnetic flux inside the green and red circles are separately shown in Fig. \ref{Fig5}(d) and (e). For the northern jet, before approximately 05:20 UT, the positive and negative magnetic flux continuously cancel with each other, and both fluxes decrease. During this period the northern jet is clearly seen, which has both base and spire structures in the first two eruptions (see Figure \ref{Fig2}), but during the third eruption it becomes very small, visible only in the SO/EUI-HRI observations as shown in Fig. \ref{Fig4} (a1–a4). The northern jet disappears when the cancellation stops as the positive polarity is nearly exhausted. Thereafter, the increase in positive flux is caused by a positive patch from the southern side moving northward. For the southern jet, the positive polarity increases and moves northward while continuously canceling with nearby negative field. Consequently, the absolute negative flux steadily decreases, whereas the positive flux first increases and later decreases. The flux decay rate is around 6 $\times$ 10$^{18}$ Mx h$^{-1}$, comparable to the findings of \citet{Chae2004} and \citet{Kaithakkal2019}, who reported similar cancellation rates in active regions. The correlation between flux cancellation and jet eruptions in our observations supports the idea that flux-cancellation–driven reconnection at the inversion line could be the primary mechanism for these recurrent jets.

\subsection{Coronal magnetic field extrapolation}

To understand the three-dimensional magnetic structure of the jets, we employed magnetic field extrapolations to infer their morphology. In the solar corona, where the plasma density is low and the plasma $\beta$ is less than unity, the magnetic field dominates the dynamics, allowing us to assume a force-free field (${\bf j} \times {\bf B} = 0$, with ${\bf B}$ the magnetic field vector and ${\bf j}$ the current density). Under the force-free assumption, the simplest case is a potential field, characterized by zero electric currents. However, in the lower solar atmosphere, where the plasma is denser and the plasma $\beta$ exceeds unity, the force-free assumption becomes invalid. In this regime, the system can be better described by a magneto-hydro-static (MHS) equilibrium, where the Lorentz force balances pressure and gravitational forces, so that one has to solve:
\begin{eqnarray}
{\bf j}\times{\bf B} & = &  \nabla p +\rho \nabla \Psi,
\label{forcebal}\\
\nabla \times {\bf B } & = &  \mu_0 {\bf j} , \label{ampere} \\
\nabla\cdot{\bf B}  & = &  0.    \label{solenoidal}
\end{eqnarray}
Here, $p$ is the plasma pressure, $\rho$ the mass density, $\mu_0$ the permeability of free space, and $\Psi$ the gravitational potential.
A recent review article on MHS is found in
\cite{Zhu2022}. For force-free fields the electric current density
is parallel to the magnetic field and the right hand side
of equation (\ref{forcebal}) vanishes. See \cite{Thomas2021}
for a review article on solar force-free fields. In its generic nonlinear form the MHS and force-free equations
require photospheric vector magnetograms as boundary condition and
are numerically expensive to solve. A special class of 
linear MHS solutions
was found by \cite{1991ApJ...370..427L} with the ansatz
\begin{equation}
\nabla \times {\bf B } = \alpha {\bf B } + a \exp(-\kappa z) \nabla B_z \times {\bf e_z}.
\label{def_j}
\end{equation}
The first term in the right-hand side of equation (\ref{def_j}) contains strictly magnetic field-aligned electric currents and contains a free parameter $\alpha$. The second term contains horizontal electric currents, which compensate the non-vanishing Lorentz force and thereby the deviation from force-freeness. Here the parameter $a$ controls the strength of these currents and the parameter $\kappa$ how fast these currents decrease with height. For the special case $(a=0)$ we get linear force-free fields and for the combination $a=0, \alpha=0$ potential fields. Linear MHS equations (including the subclasses of linear force-free and potential fields) require only the vertical photospheric magnetic field $B_z$ as boundary condition and can be solved effectively using a fast Fourier transform (FFT) method as described for potential and linear force-free fields in \citet{Alissandrakis1981} and for MHS in \citet{Thomas2015}.

\begin{figure*}[!htbp]
  \centering
  \includegraphics[trim=0cm 1.5cm 0cm 0.5cm, clip, width=0.8\textwidth]{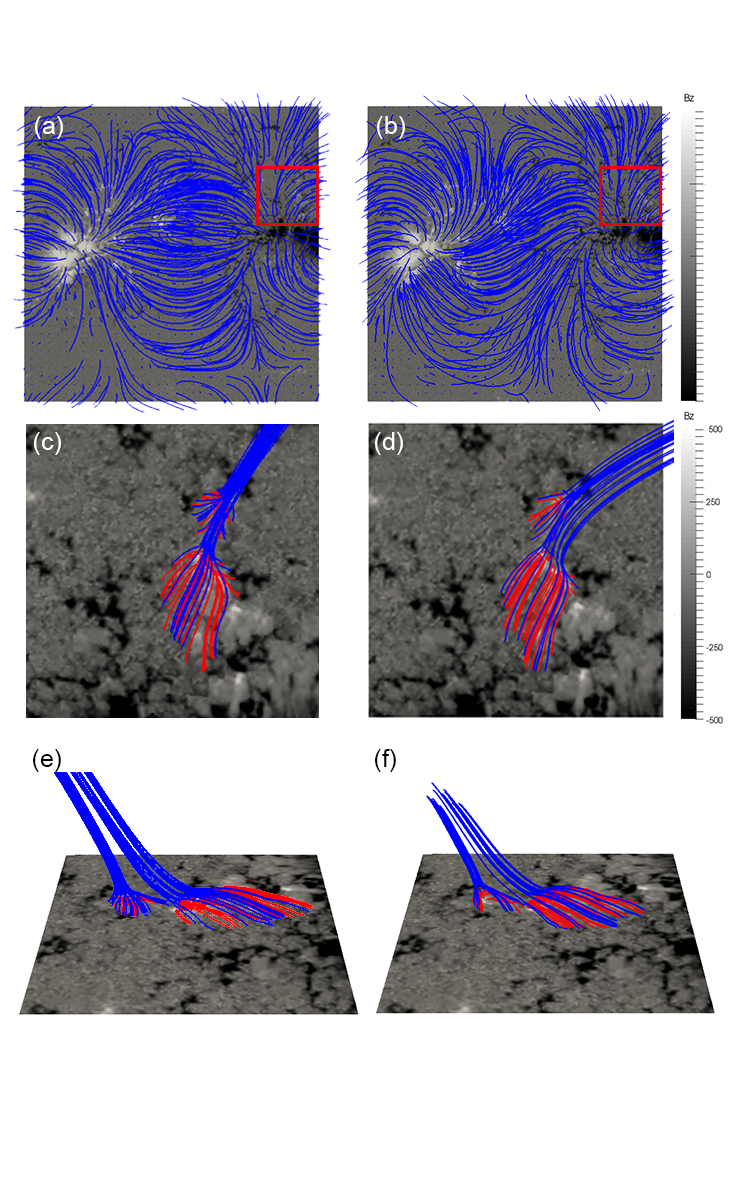}
     \caption{Potential field extrapolation (left panels) and linear MHS field extrapolation (right panels) based on the SO/PHI-HRT vector magnetic field data at 05:04:19 UT. Panels (a) and (b) are the top views with magnetic field lines marked in blue. Panels (c) and (d) show the top view of the magnetic field structures of the jets, within the FOV outlined by the red squares in the top panels. Panels (e) and (f) display the side views of the magnetic field structures of the jets. In panels (c)--(f), the blue curves represent open field lines, and the red curves represent closed field lines.}
        \label{Fig6}
  \end{figure*}

We performed both potential field extrapolation and linear MHS field extrapolation. The results are shown in Figure \ref{Fig6}. Although we also tested nonlinear force-free extrapolations, the results were found to be similar to those of the potential field, and thus are not discussed further. For the extrapolation, we need to convert the magnetic field data from the Helioprojective coordinate system of SO/PHI's detector to the Heliographic-Stonyhurst (HGS) coordinate system. We use the SO/PHI-HRT measurements of the magnetic field magnitude ($B_{\mathrm{mag}}$), inclination angle ($\theta$), and azimuth angle ($\phi$), and the method described by \citet{Gary1990} to get the field components re-projected to spherical coordinates. Then we transform the magnetic vector to the Cartesian coordinates ($B_x$, $B_y$ and $B_z$). The ambiguity of the transverse component was removed using the standard minimum‐energy disambiguation method of \citet{Metcalf1994}. Since the original SO/PHI-HRT magnetogram was not globally flux-balanced, we used the mirror magnetogram technique as developed for linear force-free fields in \citet{Seehafer1978}. The original magnetogram was extended by its three-point mirror image to ensure flux balance by construction before applying the FFT method.

For the linear MHS extrapolation we used a recently developed automatic tool \citep[see][for details]{Thomas2023, Madjarska2024} which matches individual closed magnetic field lines with small-scale loops visible in coronal images. By projecting some of the extrapolated field lines onto SO/EUI images, we optimized the linear force-free parameter to $\alpha = 4$ and the choice $a = 1.0$ and $\kappa = 0.2$ ensures a finite Lorentz force in photosphere and chromosphere and a force-free field in the corona.

Although we focused on a relatively small area of interest, namely the red box in Fig. \ref{Fig1}(e), the full SO/PHI-HRT magnetogram was used for accurate magnetic field extrapolation to minimize boundary effects. Figures \ref{Fig6}(a) and (b) present the magnetic structures of the entire AR derived from potential field and MHS extrapolations based on the full SO/PHI-HRT observation at 05:04 UT (during the third eruption). The MHS field model matches the large-scale loops visible in the SO/EUI-HRI 174 {\AA} image in Fig. \ref{Fig1}(e) more accurately, as the arcade in the lower-left region connecting the two sunspots exhibits a similar morphology. This suggests that the inclusion of pressure and gravitational forces in the MHS model provides a more realistic representation of the magnetic configuration, especially in regions with higher plasma $\beta$. The middle and bottom panels of Fig. \ref{Fig6} show the top and side views of the magnetic field structure associated with the jets inside the region marked by the red squares in \ref{Fig6}(a) and (b), providing a comprehensive 3D perspective. The northern and southern jets (indicated by the blue and red arrows in Fig. \ref{Fig1}, respectively) both exhibit fan-spine configurations: a dome-shaped fan surface separates two distinct connectivity domains, where red lines represent closed field lines and blue lines indicate open field lines. For the northern jet, a well‐defined magnetic null point can be seen in both potential and MHS models, from which the narrow spire is launched. For the southern jet, the potential field also demonstrates one null point. However, the MHS solution does not localize to a single null, instead, the null region is stretched into a sheet‐like domain, implying an extended reconnection layer.

\begin{figure*}[!htbp]
  \centering
  \includegraphics[trim=0cm 1cm 0cm 0cm, clip,  width=0.9\textwidth]{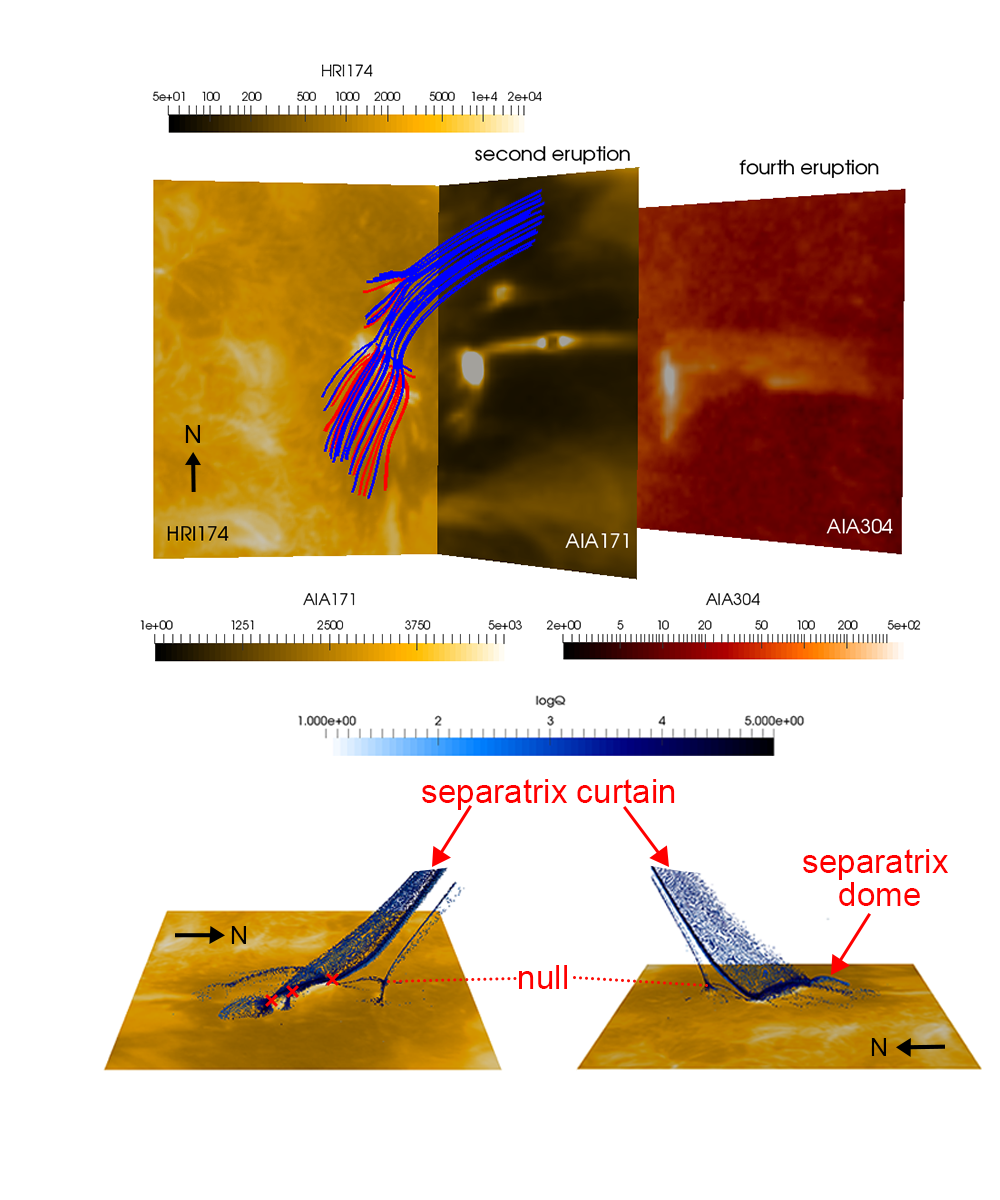}
     \caption{Linear MHS field extrapolation corresponding to Fig. \ref{Fig6}(d), overlaid on an SO/EUI‐HRI 174 {\AA} base image at 05:04:19 UT. In the upper panel, blue and red curves denote open and closed magnetic field lines, respectively. The right side displays the SDO/AIA 171 {\AA} image at 04:31:47 UT of the second jet eruption and the SDO/AIA 304 {\AA} at 05:39:41 UT of the fourth jet eruption, providing a stereoscopic perspective of the jet’s 3D geometry. The lower panels show the side views of the high-$Q$ regions ($\log Q > 5$) from the MHS extrapolation. The red crosses mark the null points associated with the southern jet.}
        \label{Fig7}
  \end{figure*}

Figure \ref{Fig7} provides a qualitative comparison between the MHS solution and the EUV observations. In the upper panel, the SO/EUI-HRI 174 {\AA} image is overlaid with open (blue) and closed (red) field lines, and two SDO/AIA images (171 {\AA} for the second eruption; 304 {\AA} for the fourth eruption) are shown at right to illustrate how the jet’s 3D structure manifests itself in observations. We can see that the domed fan surface corresponds to the bright base arch, where the internal flows are captured by the SO/EUI-HRI observations. Plasma is ejected outward from the null point along the spine, manifesting as the spire of the jet in the SDO/AIA observations. During the second eruption, the null point region is the brightest area and the spire is narrow. During the fourth eruption, the base brightening is elongated, and the spire is much wider. This elongated configuration suggests the presence of extended current sheets or complex magnetic reconnection processes. 

In order to explore further this aspect, the squashing factor \( Q \) is computed using the FastQSL method as described in \citet{ZhangP2022}. The squashing factor \( Q \) quantitatively characterizes changes in magnetic field line connectivity \citep{Titov2002, Titov2007}. Regions with high \( Q \) values delineate QSLs, which are proposed as plausible sites for magnetic reconnection \citep{Priest1995, Masson2009, Aulanier2019}. The lower panels show the high-$Q$ regions ($\log Q > 5$) from the MHS extrapolation overlaid on SO/EUI-HRI 174 {\AA} image, highlighting the separatrix surfaces that guide the plasma flows. For the northern jet, the QSL geometry matches a classic fan–spine topology involving a single null point and a dome-shaped separatrix enclosing the minority-polarity patch. The southern jet exhibits a more complex configuration with multiple null points embedded in an extended QSL system, forming a sheet-like separatrix surface extending high into the corona. This morphology closely resembles the “multi-null separatrix curtain” scenario described in Figures 8 and 9 of \citet{Pontin2022}, where reconnection at different nulls within the curtain produces multiple spatially distributed outflow channels. The observed EUI flows and the broadened AIA spire during the southern eruptions are therefore naturally explained as plasma motions confined along these high-$Q$ separatrix structures.

From the evolutionary point of view, the observed change in topology from the second to the fourth eruption can be attributed to the evolution of the magnetic field at the jet footpoint. As the positive-polarity flux patch moves northward, it continually undergoes flux cancellation with the surrounding negative-polarity field, leading to its progressive fragmentation as shown in Fig. \ref{Fig5}(c2)$-$(c4). This evolution transforms the initial single null point, characteristic of a simple fan–spine configuration, into a system containing multiple null points embedded within an extended QSL. In the MHS model, this process manifests itself as the development of a sheet-like separatrix structure (a separatrix curtain) connecting the different nulls, and reconnection at multiple topological sites along the curtain produces a broadened and more spatially distributed jet spire. Notably, true 3D perspective effects and reprojection limitations are not fully resolved in this figure, and importantly, the periodic lateral boundary conditions of our extrapolation domain force the spine to connect to the box sides, leading to the slight mismatch between the extrapolated spine and the observed jet spire.

  \begin{figure*}[!htbp]
  \centering
  \includegraphics[width=0.7\textwidth]{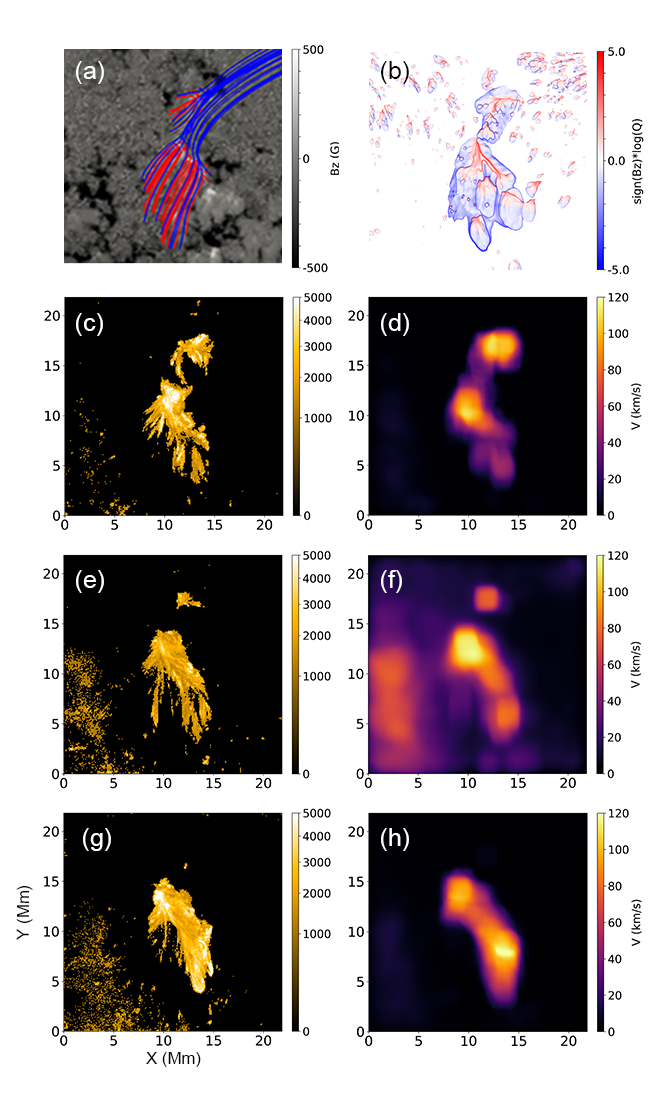}
     \caption{Panels (a) and (b) separately show magnetic structures and QSLs derived from MHS extrapolations. Panels (c), (e) and (g) display the accumulated plasma flows during the second, third and fourth eruptions, respectively. Panels (d), (f) and (h) show the corresponding mean velocity distributions.}
        \label{Fig8}
  \end{figure*}

\subsection{Velocity and temperature distribution}

In this section, we proceed to analyze the dynamics at the jet base in relation to its magnetic morphology. Panels (a) and (b) in Fig.~\ref{Fig8} present the magnetic structures and corresponding QSLs derived from linear MHS extrapolations. In panel (b), the QSL map shows the signed logarithm of the squashing factor \( Q \) computed on the photospheric plane ($z = 0$), sampled at four times the original magnetogram resolution in both directions, and masked to include only those field lines both of whose footpoints lie in the photosphere.

To examine the plasma flows associated with the jet, we use running difference maps to enhance the visibility of dynamic features. The running‐difference images are generated by subtracting each SO/EUI-HRI frame from its immediate predecessor, which enhances all moving features while suppressing the static background. An intensity threshold was then applied to each difference image to isolate only those pixels associated with the jet, ensuring that only real eruptive motions are captured. To build the integrated flow morphology, all thresholded running‐difference frames over the duration of each eruption were summed pixel–by–pixel, producing a time‐accumulated map of the jet's spatial extent, as shown in panels (c), (e), and (g) of Figure~\ref{Fig8} for the second (04:20 $-$ 04:40 UT), third (04:50 $-$ 05:20 UT), and fourth (05:30 $-$ 05:50 UT) eruptions, respectively. The north jet is still visible during the second eruption, but is hard to notice during the third and fourth eruption, consistent with our early finding that once cancellation at its footpoint has nearly stopped, only a faint northward jet, visible only in the SO/EUI-HRI observations, persists during the early phase of the third eruption.

To derive the velocity distributions, a dense optical flow technique is applied to the SO/EUI-HRI image sequences using the Farnebäck algorithm implemented in OpenCV's cv2.calcOpticalFlowFarneback() function. This technique estimates per-pixel displacements by modeling local image neighborhoods with polynomial expansions between consecutive frames \citep{Farneback2003}. The resulting vectors represent apparent motions of the jet plasma. To obtain the mean velocity map, we compute the velocity at each spatial point between all consecutive frames during the eruption period and then average the results over time. The velocity distributions corresponding to the second, third, and fourth eruptions are shown in Figure \ref{Fig8}(d), (f), and (h), respectively.

The comparison between the panels in Fig.~\ref{Fig8} reveals a clear alignment of the plasma flows with the QSLs, suggesting that the flows at the jet base are confined within the fan-shaped magnetic structure. Notably, the highest flow speeds are observed in the vicinity of the magnetic null point, indicating efficient channeling of plasma along the QSLs. During the second eruption, the highest velocities are concentrated near the magnetic null point, coinciding with the footpoints of both northern and southern jets. In the third eruption, a new null point appears below the southern fan, leading to a southward shift of the flow maximum. By the fourth eruption, strong acceleration occurs clearly in the southern region, again centered near the southern null point. This spatiotemporal evolution highlights the formation and relocation of multiple null points during the jet sequence, suggesting that successive reconnection events at distinct topological sites govern the flow dynamics. Online \texttt{Movie1.mov} further supports this observation, showing that all plasma flows originate from the null point, propagate along the fan structure. The evolution supports our previous description of the transition from a single fan-spine configuration to a more complex topology involving multiple reconnection sites. Despite this topological complexity, the observed plasma motions at the jet base remain confined within the QSLs, highlighting the controlling role of magnetic connectivity gradients in shaping the jet dynamics.

Additionally, we analyzed the temperature distribution at the base of the jet using data from the SO/SPICE instrument. Due to the unavailability of an absolute radiometric calibration for SO/SPICE, we performed a careful inspection of the wavelength bands corresponding to each spectral line. Specifically, we selected a fixed spatial pixel and inspected the time series of each spectral window to identify the number of emission components and the range over which each line core fluctuates. From these temporal stacks we determined whether a window contained one or multiple lines and estimated their approximate centroid positions and widths. We then performed Gaussian fitting and points lying within $3\sigma$ (where $\sigma$ is the standard deviation) of each preliminary centroid were flagged as the line profile. The details of the fitting methods can be found in \citet{Varesano2024}. Then, the integrated line intensity was computed by summing the area under each fitted Gaussian line. 

To obtain the temperature map, we employ the classic two-line ratio technique, using pairs of emission lines from the same element in adjacent ionization stages to serve as direct thermometers. Specifically, the O III 703.8 {\AA} / O II 718.5 {\AA} ratio sensitively tracks temperatures in the cool transition region ($4.5 < Log T[K] < 5.0$), while the O VI 1031.9 {\AA} / O V 760.4 {\AA} ratio probes the warmer upper transition region ($5.3 < Log T[K] < 5.6$). Under the spectral resolution of SPICE, both the O III 703.8 {\AA} and O II 718.5 {\AA} features consist of unresolved blends of multiple nearby emission lines. To account for this, we calculated the theoretical contribution function $G(T)$ of each line by summing the contributions of all transitions falling within ±0.195 {\AA} of the target wavelength, corresponding to the spectral pixel size as recorded in the FITS header. This correction ensures that our modeled line ratios accurately reflect the observed features. By computing each line's contribution function \(G(T)\) with CHIANTI \citep{DelZ2015} with an assumed electron density of $n_e= 2 \times 10^9 cm^{-3}$, forming the theoretical ratio curve
\[
R_{\rm th}(T) \;=\; \frac{G_{\rm higher}(T)}{G_{\rm lower}(T)}\,,
\]
and then inverting our observed SO/SPICE intensity ratios via interpolation, we generate temperature maps that are independent of elemental abundance and only weakly dependent on density. 

\begin{figure*}[!htbp]
  \centering
  \includegraphics[width=\textwidth]{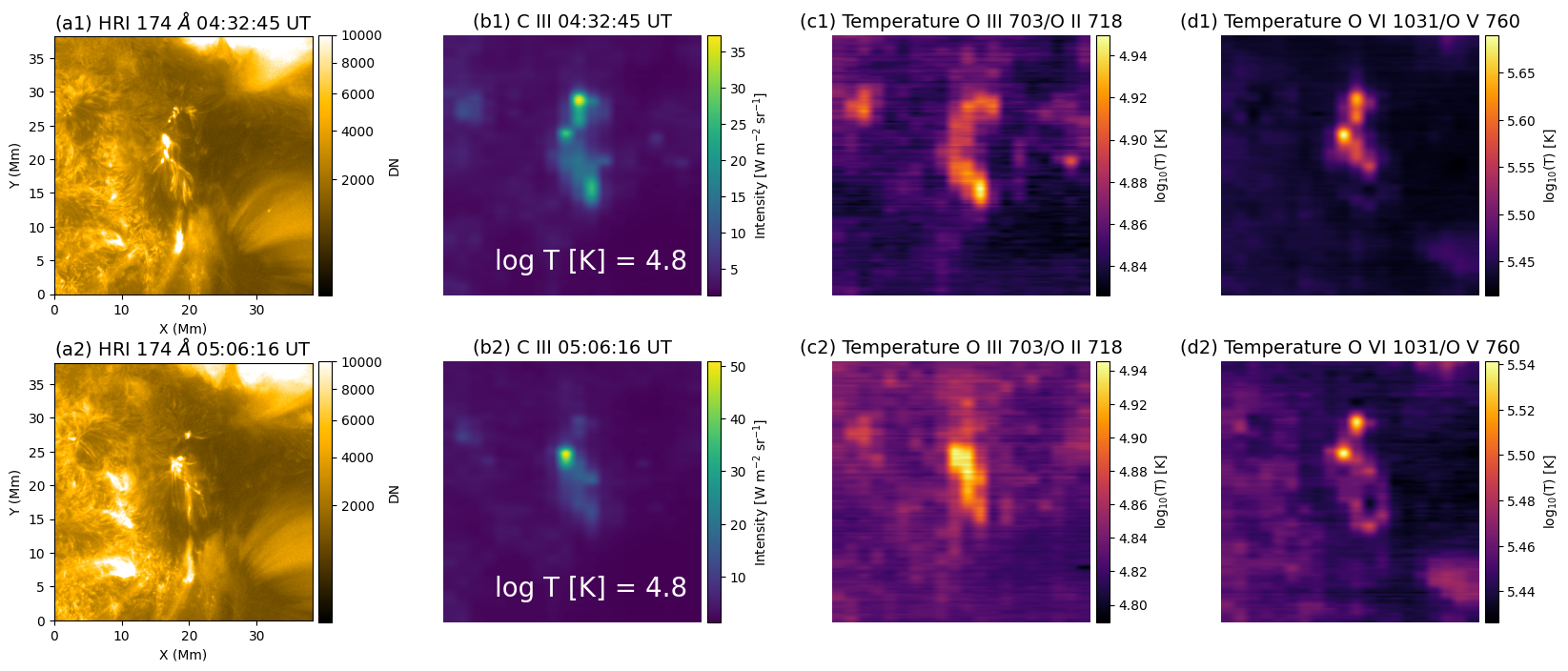}
     \caption{Co‐aligned SO/EUI‐HRI 174 {\AA} images (first column), SO/SPICE C III 976 {\AA} line–integrated intensity maps (second column), and temperature maps derived from the O III 703.8 {\AA} / O II 718.5 {\AA} ratio (third column) and the O VI 1031.9 {\AA} / O V 760.4 {\AA} ratio (fourth column). The top row shows the second eruption of the recurrent jet, and the bottom row shows the third eruption.}
        \label{Fig9}
  \end{figure*}

The electron density of $n_e= 2 \times 10^9 cm^{-3}$ is consistent with typical values reported in previous SO/SPICE studies \citep{Varesano2024, Brooks2024}. To assess the robustness of our diagnostics, we also tested a broader density range from $10^9 cm^{-3}$ to $5 \times 10^{10} cm^{-3}$, and found that the resulting line ratio curves and derived temperature maps showed negligible variation across this range. Figure \ref{Fig9} displays the SO/EUI-HRI 174 {\AA} image, the integrated intensities of the C III (977.03 {\AA}) and the temperature maps inferred from O III 703.8 {\AA} / O II 718.5 {\AA} and O VI 1031.9 {\AA} / O V 760.4 {\AA} during the second eruption (upper row) and the third eruption (lower row). The structure of the jet base can be easily distinguished in the panels (c1) and (c2). Although we do not show all the SO/SPICE observations here, the jet base exhibits a similar response across a broad range of temperatures, indicating that it consists of multi-thermal structures. Notably, the region near the magnetic null point displays higher temperatures, as shown in the O VI 1031.9 {\AA} / O V 760.4 {\AA} temperature maps, suggesting that this is the site where magnetic reconnection is occurring.

\section{Discussion and Conclusions}  \label{sec:con}
Solar Orbiter’s stereoscopic perspectives, unprecedented spatial resolution, and high‐cadence imaging, together with SDO observations, enable us to conduct a continuous four-hour study of recurrent jets and to reconstruct their 3D dynamics in detail. The jet spire exhibits a multi-thermal structure with apparent velocities around 100 km s$^{-1}$, as observed by SDO/AIA. By combining high-resolution imaging from SO/EUI-HRI with magnetic field extrapolations based on SO/PHI-HRT data, we found these jets to exhibit a fan-spine magnetic topology. Over four eruptions, the magnetic topology evolves from a simple fan–spine structure with a single null point to a more complex system containing multiple null points with an extended separatrix curtain. This topological evolution is accompanied by a morphological transition of the jets, from narrow, well-collimated spires to broader, more fragmented outflows with multiple bright strands, reflecting the formation of multiple reconnection channels. The flows at the base are confined within the dome-shaped QSLs, with the highest velocities and temperatures occurring near the null point.

\subsection{Recurrent jets and magnetic reconnection}
The recurrent nature of the jets, as revealed by the time-distance plots, suggests that they are triggered by a repetitive reconnection process at the magnetic null point. This is in agreement with a previous study by \citet{Cheng2023}, proving that the null point is persistent. The observed magnetic flux cancellation rate of approximately 6 $\times$ 10$^{18}$ Mx h$^{-1}$ over a period of 2.5 hours is consistent with previous studies \citep{Chae2004, Kaithakkal2019}, indicating that continuous flux cancellation at the magnetic inversion line could sustain the repetitive reconnection process. This finding supports the scenario proposed by \citet{Panesar2016}, where magnetic reconnection driven by flux cancellation is responsible for triggering jet eruptions.

The co-alignment of SO/SPICE data with SO/EUI-HRI 174 {\AA} images further revealed a multi-thermal response at the jet base, with higher temperatures observed near the magnetic null point. This is in agreement with previous observations that solar jets often contain multi-thermal plasma components \citep{Nistico2009, Sterling2015}. The multi-thermal nature of the jets implies that different layers of the solar atmosphere are involved in the reconnection process, likely through a combination of heating and direct ejection of pre-existing cool plasma. Numerical simulations \citep{Johnson2024} indicate a higher temperature near the null point. This high temperature can be attributed to several factors: besides the direct consequence of heating due to magnetic reconnection, where magnetic energy is rapidly converted into thermal energy \citep{Craig1991, McLaughlin2006}, magnetic field stress near null points can lead to strong current accumulation, which dissipates through Ohmic heating, warming a significant plasma volume \citep{Galsgaard2011}. Also, as the magnetic field weakens around null points, thermal conduction transitions from anisotropic to isotropic, enhancing local heating efficiency \citep{Braginskii1965}. Additionally, the highly non-uniform magnetic field can inhibit conductive cooling due to strongly diverging field lines, causing heat to accumulate \citep{Antiochos1976, Cargill2022}. 

\subsection{Fan-spine magnetic topology and its evolution}

Building on the reconnection scenario above, our MHS extrapolations prove that these null‐point reconnections occur within a fan–spine configuration, characterized by a dome-shaped fan surface that separates different connectivity domains, with closed and open magnetic field lines connecting at a null point. The presence of fan-spine structures in both the northern and southern jets suggests that this configuration is a common feature in recurrent jet phenomena, as in most previous studies recurrent jets also show a fan-spine structure \citep{Duan2024, Joshi2024, Zhou2025}. The alignment of plasma flows with the QSLs derived from the MHS extrapolation indicates that the observed plasma dynamics are strongly guided by the magnetic topology.

\citet{Pontin2022} summarize how magnetic topology as described by nulls, separatrices, and QSLs governs current-sheet formation and distinct 3D reconnection regimes in the corona, and demonstrate that a single null embedded in opposite-polarity flux forms a dome-shaped fan with a narrow outer spine, whereas multiple nulls connected by separators produce a curtain-like separatrix sheet spanning a broader angular sector. Our high‐resolution SO/EUI and SDO observations, together with MHS extrapolations, capture this transition in action and link it directly to the photospheric evolution on one side, and to changes in jet morphology on the other side. As the positive‐polarity flux at the jet base drifts northward and fragments through ongoing cancellation with surrounding negative flux, the initial single‐null dome evolves into a cluster of nulls connected along an extended separatrix footprint, forming a curtain‐like topology in the extrapolation. Morphologically, this corresponds to a clear progression from narrow, well‐collimated spires (first two eruptions) to much broader, fragmented outflows with multiple bright strands (fourth eruption). The SO/EUI-HRI 174~{\AA} images show the base brightening stretching and splitting into several luminous cores, while SDO/AIA 304~{\AA} reveals the spire widening accordingly. Such a transition naturally facilitates the formation of extended current sheets, enabling more efficient reconnection and generating multi-stranded, multi-thermal outflows \citep{Masson2009, Wyper2017, Joshi2024}. These results provide direct observational evidence for dynamic null-point bifurcation modulating jet morphology and energetics in small-scale recurrent events.

\subsection{Future prospects}
In this work, we combine unique multi‐viewpoint, multi‐line observations with state-of-the-art modeling techniques to provide a comprehensive 3D picture of recurrent jets, including their magnetic topology, eruption mechanism, evolution and temperature structures, which allow a coherent interpretation of their observational signatures. In particular, our study highlights the importance of fan-spine magnetic structures and magnetic reconnection at null points in driving recurrent solar jets. Moreover, the Solar Orbiter’s stereoscopic vantage provides unprecedented 3D insight into the formation and dynamics of these jets, substantially advancing our understanding of their underlying physical processes.

Future studies could benefit from higher temporal and spatial resolution observations to better resolve fine-scale reconnection dynamics and the structure of extended null regions. Moreover, performing more advanced extrapolation techniques, such as nonlinear force-free field (NLFFF) or nonlinear MHS models, could provide a more realistic insight into the complexity of the magnetic field configurations associated with solar jets. Additionally, numerical simulations that incorporate the observed fan-spine topology and multi-thermal plasma characteristics could help in quantifying the role of magnetic reconnection and testing the robustness of the observed correlations.

\begin{acknowledgements}
We thank Solar Orbiter and SDO science teams for the valuable data. Solar Orbiter is a space mission of international collaboration between ESA and NASA, operated by ESA. We are grateful to the ESA SOC and MOC teams for their support. The German contribution to SO/PHI is funded by the BMWi through DLR and by MPG central funds. The Spanish contribution is funded by AEI/MCIN/10.13039/501100011033/ and European Union ``NextGenerationEU/PRT'' (RTI2018-096886-C5, PID2021-125325OB-C5, PCI2022-135009-2, PCI2022-135029-2) and ERDF ``A way of making Europ''; ``Center of Excellence Severo Ocho'' awards to IAA-CSIC (SEV-2017-0709, CEX2021-001131-S); and a Ramóny Cajal fellowship awarded to DOS. The French contribution is funded by CNES. This project has received funding from the European Research Council (ERC) under the European Union's Horizon 2020 research and innovation programme (grant agreement No. 101097844 — project WINSUN). TW acknowledges DLR grant 50OC2301.
\end{acknowledgements}

\bibliographystyle{aa} 
\bibliography{jet} 

\end{document}